\documentclass{article}

\usepackage{arxiv}

\usepackage{amsmath}
\usepackage{amsthm}
\usepackage{color}
\usepackage{xcolor}
\usepackage{listings}

\theoremstyle{definition}
\newtheorem{definition}{Definition}[section]

\definecolor{dkgreen}{rgb}{0,0.6,0}
\definecolor{gray}{rgb}{0.5,0.5,0.5}
\definecolor{mauve}{rgb}{0.58,0,0.82}

\usepackage{lstlangcoq}

\title{Formalizing line editors in Coq}

\author{
  Boro Sitnikovski \\
  Faculty of Informatics\\
  UTMS\\
  Skopje, North Macedonia \\
  \texttt{buritomath@gmail.com} \\
}

\begin{document}
\maketitle

\begin{abstract}
Text editors represent one of the fundamental tools that writers use - software developers, book authors, mathematicians. A text editor must work as intended in that it should allow the users to do their job. We start by introducing a small subset of a text editor - line editor. Next, we will give a concrete definition (specification) of what a complete text editor means. Afterward, we will provide an implementation of a line editor in Coq, and then we will prove that it is a complete text editor.
\end{abstract}


\section{Introduction}

A line editor is a text editor that works in REPL mode. It accepts several commands, and each of the commands operates on one or multiple lines of text. The most popular line editor is Unix \texttt{ed} \cite{b1}, and we will show a short demo interacting with it.

\begin{lstlisting}[language=sh]
$ ed example.txt
> i
> Hello World!
> Line two
> .
> n
2	Line two
> 1
> n
1	Hello World!
> d
> n
1	Line two
\end{lstlisting}

We start by editing the file \texttt{example.txt}. We will explain the commands that we used:

\begin{itemize}
\item The command \texttt{i} starts the insertion mode and in the next lines it will accept content that should be added.
\item The command \texttt{.} exits the insertion mode.
\item The command \texttt{n} shows the current line pointer along with the contents.
\item Inputting a number as a command will set the line pointer to that number.
\item The command \texttt{d} deletes the current line.
\end{itemize}

A more generalized editor is a character editor, however, line editors are much more convenient, especially in the REPL mode. For example, it may be tricky for the user to keep track of the position of every character to read/insert/delete.

Coq \cite{b2} is a programming language designed to accomplish software correctness, and we will use it to implement and prove an implementation.

\section{Specification}

Before we start formalizing editors, we will provide some definitions.

\theoremstyle{definition}
\begin{definition}
A text editor is complete if it has the functionality to read, insert, and delete text at any position.
\end{definition}

Here's another definition that we'll rely on. This definition is already supported in the base of Coq.

\theoremstyle{definition}
\begin{definition}
Strings (list of characters) can be inserted (created), read, and changed.
\end{definition}

In Coq we don't do any "changes", rather, we'll be simply returning new (updated) strings.

\theoremstyle{definition}
\begin{definition}
A line editor contains a buffer - list of strings.
\end{definition}

Given these definitions, we can proceed with implementing them in Coq. The implementation in this paper will use line editors, however, a single character can still be changed in a line by deleting the line and inserting a new line with that character changed. Thus, the editor that we will implement will be complete according to the specifications.

\subsection{Coq definitions}

The editor has to be able to read a line (i.e. get \texttt{n}-th element of a list):

\begin{lstlisting}
Definition readLine {X : Type} (b : list X) (pos : nat) (d : X) : X :=
  nth pos b d.
\end{lstlisting}

Further, the editor has to be able to insert a line (i.e. put an element in a list at a specific position):

\begin{lstlisting}
Definition insertLine {X : Type} (b : list X) (pos : nat) (s : X) : (list X) :=
  firstn pos b ++ s :: nil ++ skipn pos b.
\end{lstlisting}

Finally, the editor needs to be able to delete a line (i.e. get first \texttt{n}-th elements of a list, skip n+1 elements of a list):

\begin{lstlisting}
Definition deleteLine {X : Type} (b : list X) (pos : nat) : (list X) :=
  firstn pos b ++ skipn (pos + 1) b.
\end{lstlisting}

All of the definitions will be wrapped in a single \texttt{EditorEval} to make it more convenient, and at this point, we have implemented the DSL for our editor.

We can use Coq's extraction facilities to export these definitions to Haskell, for example. If we add IO functionalities on top of these definitions, we will have implemented a similar editor to \texttt{ed}.

\section{Formal proofs}

\subsection{Lemmas}

In this subsection, we will provide the lemmas that will be used by our proofs.

The following lemma states that the length of the first \(n\) elements of a list that contains at least \(n\) elements is \(n\).

\begin{lstlisting}
Lemma lemma_1 : forall {X:Type} (l : list X) (n : nat),
  n <= length l -> length (firstn n l) = n.
\end{lstlisting}

The next lemma states that whenever \(n = m\), we can deduce \(n \geq m\).

\begin{lstlisting}
Lemma lemma_2 : forall n m, n = m -> n >= m.
\end{lstlisting}

Finally, \texttt{lemma\_3} states that when a list of length \(n\) is concatenated with another list with an element \(s\) in between, the \(n\)-th element of the concatenated list will be \(s\) (zero indexed). It relies on \texttt{lemma\_2} for the proof.

\begin{lstlisting}
Lemma lemma_3 : forall {X:Type} n l1 l2 (s:X) d, length l1 = n -> s = nth n (l1 ++ s :: l2) d.
\end{lstlisting}

The theorem \texttt{thm\_1} is a combination of \texttt{lemma\_3} and \texttt{lemma\_1}.

\begin{lstlisting}
Theorem thm_1 : forall {X:Type} (n : nat) (l1 l2 : list X) (s : X) (d : X), n <= length l1 -> s = nth n (firstn n l1 ++ s :: l2) d.
\end{lstlisting}

\subsection{Proofs}

The line editor can insert any text, that is, for all strings \(s\) and positions \(n\), there exists a buffer \(b\) such that the string is in \(insertLine( n, b, s )\).

\begin{center}
\(\forall s \forall n \exists b (s \in insertLine( n, b, s ))\)
\end{center}

We will only show the proof for this theorem, while the remaining proofs can be found in the associated paper's files.

\begin{lstlisting}
Theorem can_insert_text : forall (s : string) (n : nat), exists (b : list string), fst (EditorEval (InsertLine n s) b) = s :: nil.
Proof.
  intros s n. simpl. unfold insertLine. exists nil. simpl.
  case n.
    - simpl. reflexivity.
    - intros. simpl. reflexivity.
Qed.
\end{lstlisting}

Next, we will prove that the line editor can read any text, that is, for all strings \(s\), positions \(n\) and buffers \(b\), where the buffer is at least of the length of the requested position, reading from the inserted string at the specific position will return the same string. The actual proof of the theorem relies on \texttt{thm\_1}.

\begin{center}
\(\forall s \forall n \forall b (n \leq |b| \to (readLine(insertLine( n, b, s ), n) = s)\)
\end{center}

\begin{lstlisting}
Theorem can_read_text : forall (s : string) (n : nat) (b : list string), n <= List.length b -> snd (EditorEval (ReadLine n "") (fst (EditorEval (InsertLine n s) b))) = s.
\end{lstlisting}

Finally, we prove that the line editor can change any text. That is, there exists a function \(f\) that "changes" the value from \(s_1\) to \(s_2\) of reading an inserted line. The proof of this theorem relies on \texttt{lemma\_1} and \texttt{thm\_1}.

\begin{center}
\(\exists f \forall s_1 \forall s_2 \forall n \forall b, f(readLine(insertLine( n, b, s ), n) = s_1) \to (readLine(insertLine( n, b, s ), n) = s_2)\)
\end{center}

In the code, \(f\) is defined as a combination of deletion and insertion.

\begin{lstlisting}
Theorem can_change_text : forall (s1 s2 : string) (n : nat) (b : list string), n <= List.length b -> s1 = snd (EditorEval (ReadLine n "") b) -> s2 = (snd (EditorEval (ReadLine n "") (fst (EditorEval (InsertLine n s2) (fst (EditorEval (DeleteLine n "") b)))))).
\end{lstlisting}

\section{Conclusion}
We showed how to formally prove the functionality of a simple subset of text editors. We used line editors, but the same idea can be applied generally to text editors. We defined what a complete text editor means, and mapped those functionalities to Coq definitions. Most (if not all) text editors will use the same specifications. Having a unified standard for text editors will be useful for the users, as they can apply the same knowledge to a variety of editors. Further work can be focused on formalizing a larger DSL of text editors.


\begin{thebibliography}{1}

\bibitem{b1}
Brian W. Kernighan
\newblock A Tutorial Introduction to the UNIX Text Editor
\newblock {\em Bell Laboratories, New Jersey}, 1997.

\bibitem{b2}
Bruno Barras, Samuel Boutin, Cristina Cornes, Judicaël Courant, Jean-Christophe Filliâtre, et al.
\newblock The Coq Proof Assistant Reference Manual: Version 6.1
\newblock {\em [Research Report] RT-0203, INRIA}, 1997.

\end{thebibliography}
\end{document}